\documentclass[]{copernicus}

\begin{document}

\title{Turbulent spectra and spectral kinks in the transition range from MHD
to kinetic Alfv\'{e}n turbulence}

\author[]{Y. Voitenko}
\author[]{J. De Keyser}

\affil[]{Belgian Institute for Space Aeronomy, Brussels, Belgium}


\runningtitle{MHD/kinetic turbulence transition}

\runningauthor{Y. Voitenko and J. De Keyser}

\correspondence{Y. Voitenko\\ (voitenko AT oma.be)}

\received{}
\pubdiscuss{} 
\revised{}
\accepted{}
\published{}


\firstpage{1}

\maketitle

\begin{abstract}
A weakly dispersive sub-range (WDR) of kinetic Alfv\'{e}n turbulence is
distinguished and investigated for the first time in the context of
MHD/kinetic turbulence transition. We found perpendicular wavenumber spectra
$\propto k_{\bot }^{-3}$ and $\propto k_{\bot }^{-4}$ formed in WDR by
strong and weak turbulence of kinetic Alfv\'{e}n waves (KAWs), respectively.
These steep WDR spectra connect shallower spectra in the MHD and strongly
dispersive KAW sub-ranges, which results in a specific double-kink (2-k)
pattern often seen in observed turbulent spectra. The first kink occurs
where MHD turbulence transforms into weakly dispersive KAW turbulence;
the second one is between weakly and strongly dispersive KAW sub-ranges. Our
analysis suggests that the partial turbulence dissipation due to
amplitude-dependent super-adiabatic ion heating may occur in the vicinity of
the first spectral kink. A threshold-like nature of this process results in
a conditional selective dissipation affecting only largest over-threshold
amplitudes and decreasing intermittency in the range below the first spectral kink.
Several recent counter-intuitive observational findings can be explained by the
selective dissipation coupled to the nonlinear interaction among weakly
dispersive KAWs.
\end{abstract}

\introduction

Kinetic Alfv\'{e}n waves (KAWs) are an extension of MHD Alfv\'{e}n waves in
the range of high perpendicular wavenumbers $k_{\perp }$ in the plane $\perp
\mathbf{B}_{0}$, where linear and nonlinear effects due to finite values of $%
\mu =k_{\perp }\rho _{i}$ become significant ($\mathbf{B}_{0}\parallel z$ is
the background magnetic field, $\rho _{i}=V_{Ti}/\Omega _{i}$ is the ion
gyroradius) (Hasegawa and Chen, 1976; Voitenko, 1998a). There are numerous
observational and theoretical indications that MHD Alfv\'{e}n turbulence in
the solar wind cascades towards high $k_{\perp }$ and eventually reaches the
KAW wavenumber range at the proton gyroradius scales, $k_{\perp }\rho
_{p}\sim 1$ (Leamon et al., 1999; Bale et al., 2005; Alexandrova et al.,
2008; Sahraoui et al., 2010). It is not yet certain what happens next with
these KAWs: do they dissipate heating plasma (Leamon et al. 1999), or
interact nonlinearly among themselves and proceed cascading further towards
higher $k_{\perp }$, reaching electron scales (Alexandrova et al., 2008;
Sahraoui et al., 2010). It has been envisaged that the nonlinear evolution
and related wavenumber spectra in the range below the spectral break point $%
k_{\perp b}\sim \rho _{p}^{-1}$ are dominated by MHD-type nonlinear
interactions among Alfv\'{e}n waves, and the spectra at $k_{\perp }>k_{\perp
b}$ are determined by linear and nonlinear properties of KAWs. If the
dissipation is more efficient, the cascade should dissipate in the vicinity
of $k_{\perp b}$ and cannot reach electron scales, as was argued by Leamon
et al. (1999), Howes et al. (2008), and Podesta et al. (2009) using Landau
damping estimations. There are, however, observational indications (see
Sahraoui et al., 2010, and references therein), that the nonlinear
interaction among KAWs is faster than their dissipation, and the turbulence
cascade proceeds further at higher $k_{\perp }\rho _{p}\gg 1$ creating a
kinetic-scale turbulence of KAWs.

From observational point of view, the transformation occurs at the spectral
break points $f_{b}$ dividing shallower MHD spectra $\propto f^{-5/3}$ at $%
f<f_{b}$ and steeper kinetic spectra with power indexes ranging from $-2$ to
$-4$ at $f<f_{b}$, which are observed in the solar wind ($f$ is the
frequency in the satellite frame). Because of the large solar wind velocity,
$V_{SW}\gg V_{A}$, the Alfv\'{e}nic time variations $\omega _{A}\sim
k_{z}V_{A}$ are usually much slower than the Doppler frequencies in
satellite frame $\omega _{d}=\mathbf{k}\cdot \mathbf{V}_{SW}$ (Tailor
hypothesis). Then the satellite-frame frequency spectra are dominated by the
Doppler frequency, $2\pi f=$ $\left\vert k_{z}V_{A}-\mathbf{k}\cdot \mathbf{V%
}_{SW}\right\vert $ $\sim \left\vert \mathbf{k}\cdot \mathbf{V}%
_{SW}\right\vert $, representing wave-number spectra. As the solar wind
turbulence is dominated by large perpendicular wave vectors $k_{\perp }\gg
k_{z}$ (Sahraoui et al., 2010; Luo and Wu, 2010, and references therein),
satellites measure perpendicular wavenumber spectra, $f\propto k_{\perp }$.
There are of cause rare cases $\mathbf{B}_{0}\parallel \mathbf{V}_{SW}$,
where frequency measures parallel wavenumber, $f\propto k_{z}$. The spectral
break $f_{b}$ of the turbulence is often associated with one of the proton
kinetic scales, proton gyroradius $\rho _{p}$ or proton inertial length $%
\delta _{p}=V_{A}/\Omega _{i}$, such that the observed break-point frequency
$2\pi f_{b}\simeq $ $V_{SW}/\rho _{p}$ or $V_{SW}/\delta _{p}$ (Leamon et
al., 1999; Bale et al., 2005; Alexandrova et al., 2008; Sahraoui et al.,
2010).

Because of the complex interplay between linear and nonlinear dynamics of
KAWs, theoretical interpretation of the turbulence, its dissipation, and
related spectra in the KAW range is still incomplete. In particular, recent
theoretical analysis by Podesta et al. (2010) argues that the KAW cascade
subject to collisionless Landau damping cannot reach electron scales in the
solar wind conditions, which contradicts the opposite conclusion by Sahraoui
et al. (2010) based on observations. Using Cluster data, Sahraoui et al.
(2010) found that the wavenumber spectra of MHD and KAW turbulences have
slopes $\propto k_{\perp }^{-1.7}$ and $\propto k_{\perp }^{-2.8}$,
respectively, and the KAW turbulence extends to electron scales in the solar
wind at 1 AU (Sahraoui et al., 2010). Between these $k_{\perp }^{-1.7}$ and $%
k_{\perp }^{-2.8}$ spectra, Sahraoui et al. (2010) also noticed much steeper
$\propto k_{\perp }^{-4}$ spectra that appear in the weakly/mildy dispersive
KAW sub-range $0.6<k_{\perp }\rho _{p}<2$ (see their Fig. 1). The same
spectral form in the MHD/kinetic transition range, containing two spectral
kinks with steepest spectra inbetween, can be seen in other recent studies -
see example in Fig.1 adopted from the paper by Chen et al. (2010) (see also
Fig. 1 by Smith et al., 2006).

\begin{figure}[]
\vspace*{2mm}
\par
\begin{center}
\includegraphics[width=8.3cm]{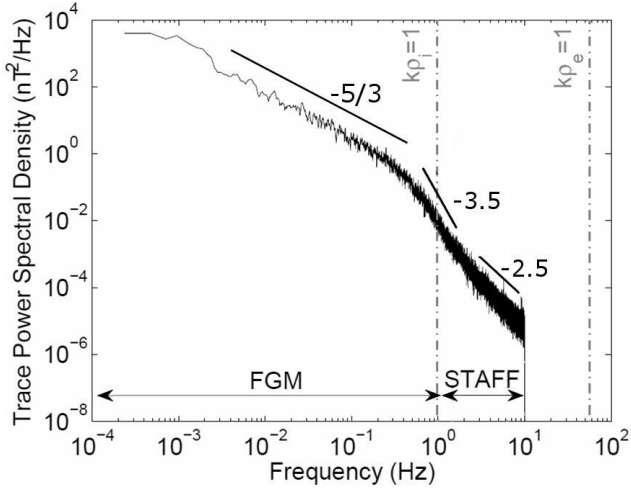}
\end{center}
\caption{Turbulent Alfv\'{e}nic spectrum extending over three sub-ranges, with the
steepest slope in the weakly/mildy dispersive KAW sub-range (interpretation
of the spectrum shown in Fig. 1 by Chen et al., 2010)}
\label{fig1}
\end{figure}

Steep variable spectra in the same wavenumber range were observed before
(Leamon et al. 1999), but without connection to shallower higher-wavenumber
(higher-frequency in the satellite frame) spectra that were unavailable.
These steep spectra were called the "dissipation range" spectra, and were
associated with dissipation, mainly via kinetic ion-cyclotron and Landau
damping. However, the nature of the "dissipation range" and its spectra is
not so clear. For example, recent observations of reduced magnetic helicity
imply the presence of counter-streaming KAWs surviving the "dissipation
range" rather than ion-cyclotron damping in it (Carbone et al. 2010).

Analyzing ACE spacecraft data, Smith et al. (2006) have found that the
larger spectral fluxes (as measured at 0.01 Hz) are followed by the steeper
spectra in the "dissipation range" above the spectral break $f_{b}\sim 0.3$
Hz. This counterintuitive observational fact is difficult to explain by
ion-cyclotron and Landau damping. Smith et al. (2006) did not find any
regular dependence of $f_{b}$ on the cascade rate. However, later on
Markovskii, Vasquez, and Smith (2008) studied the statistics and scaling of
the spectral breaks and concluded that their positions are determined by a
combination of their scales and the turbulent amplitudes at that scales,
which suggests a non-linear dissipation mechanism for the solar wind
turbulence. Again, kinetic ion-cyclotron and Landau damping mechanisms would
not lead to such behavior of the dissipation range.

Motivated by these findings, in the present paper we analyze the transition
wavenumber range from MHD to KAW turbulence focusing on nonlinear KAW
properties. We demonstrate that the observed spectral forms and steep
spectra in the "dissipation range" can be explained by the nonlinear
interaction of weakly dispersive KAWs without involving kinetic
ion-cyclotron and Landau dissipation mechanisms.

\section{Weakly dispersive $k_{\perp }\protect\rho _{p}<1$ sub-range of the
KAW turbulence}

In the weakly dispersive range, in the limit $k_{\perp }^{2}\rho _{p}^{2}\ll
1$, the rate of the nonlinear interaction among highly oblique $k_{z}\ll
k_{\perp }$ co-propagating KAWs is calculated by maximizing the matrix
element of the 3-wave KAW interaction calculated by Voitenko (1998a,b):
\begin{equation}
\gamma _{k\uparrow \uparrow }^{NL}\simeq 0.4\Omega _{p}\frac{V_{A}}{V_{Tp}}%
\mu ^{3}\frac{B_{k}}{B_{0}},  \label{g.co.WD}
\end{equation}%
where $B_{k}$ is the KAW amplitude at the (anisotropic) length scales $%
\lambda _{z}=2\pi /k_{z};$ $\lambda _{\perp }=2\pi /k_{\perp }$, where $%
k_{z} $ and $k_{\perp }$ are parallel and perpendicular KAW wavenumbers. We
put $T_{e\parallel }/T_{p\perp }=1$ in (\ref{g.co.WD}) and further on for
simplicity.

In the KAW case $k_{\perp }\gg k_{z}$ and the nonlinear wave dynamics is
driven by the energy exchange among short cross-field length scales, whereas
the parallel scales follow the perpendicular ones kinematically (in the weak
turbulence) or via critical balance (in the strong turbulence). Then the
amplitude $B_{k}$ can be related to the omnidirectional spectral energy
density $W_{k\perp }$ as $B_{k}=\sqrt{k_{\perp }W_{k\perp }}$ ($W_{k\perp }$
is defined such that $\int_{0}^{\infty }dk_{\perp }W_{k\perp }=$ total
fluctuation energy per unit volume).

The nonlinear interaction rate of counter-propagating $\mu \ll 1$ KAWs is
(Voitenko, 1998a,b)%
\begin{equation}
\gamma _{k\uparrow \downarrow }^{NL}\simeq 0.3\Omega _{i}\frac{V_{A}}{V_{Ti}}%
\mu ^{2}\frac{B_{k}}{B_{0}}.  \label{g.con.WD}
\end{equation}%
In the weakly dispersive sub-range $\gamma _{k\uparrow \uparrow
}^{NL}<\gamma _{k\uparrow \downarrow }^{NL}$, but short linear correlation
time among counter-propagating KAWs can reduce the counter-propagating
nonlinear rate below $\gamma _{k\uparrow \uparrow }^{NL}$. Note that there
is no explicit $k_{z}$- dependence in the above expressions for interaction
rates.

\subsection{Weak KAW turbulence}

The nonlinear interaction among co-propagating KAWs can be considered weak
if their nonlinear rate (\ref{g.co.WD}) is less than the dispersive part of
the KAW frequency: $\gamma _{k}^{NL}<\delta \omega _{k}$, where $\delta
\omega _{k}\simeq k_{z}V_{A}\mu ^{2}$ in isothermal plasmas. In this case,
the conservation law for the generalized enstrophy (dispersive part of
energy) apply, and nonlinear interaction among co-propagating KAWs (\ref%
{g.co.WD}) establishes perpendicular wavenumber spectra
\begin{equation}
W_{k}\propto k_{\perp }^{-5};  \label{2D.co.WD.WT1}
\end{equation}%
\begin{equation}
W_{k}\propto k_{\perp }^{-4},  \label{2D.co.WD.WT2}
\end{equation}%
created by the direct enstrophy and inverse energy cascades, respectively
(Voitenko, 1998b).

For the axially symmetric turbulence in the cross-field plane we can define
a reduced omnidirectional spectral power $W_{k\perp }$ $=2\pi k_{\perp }\int
dk_{z}B_{k}^{2}$, such that $W=\int dk_{\perp }W_{k\perp }$. The energy
exchange among different $k_{\perp }$ does not depend on $k_{z}$ (cf. eqs
(6.1) and (6.2) by Voitenko (1998a)). Hence the weakly turbulent 3D spectra (%
\ref{2D.co.WD.WT1}-\ref{2D.co.WD.WT2}) have their corresponding reduced
omnidirectional power spectra
\begin{equation}
W_{k\perp }\propto k_{\perp }^{-4},  \label{1D.co.WD.WT1}
\end{equation}%
\begin{equation}
W_{k\perp }\propto k_{\perp }^{-3}.  \label{1D.co.WD.WT2}
\end{equation}%
The omnidirectional wavenumber spectra are those measured in the solar wind
by satellites as 1D Doppler frequency spectra if the solar wind velocity $%
\mathbf{V}_{SW}\nparallel \mathbf{B}_{0}$. In the case of turbulent spectra
that are axially asymmetric around $B_{0}$ axis, the measured 1D spectrum
may have a larger power index, approaching $W_{k\perp }\propto k_{\perp
}^{-5}$ in the extreme case of flat turbulence (say, $\propto k_{x}^{-5}$).
Accounting for a possible anisotropy, the steepest spectra produced by the
weakly dispersive turbulence are
\begin{equation}
W_{k\perp }\propto k_{\perp }^{-4}\div k_{\perp }^{-4.5}.
\label{1D.co.WD.WT3}
\end{equation}

Nonlinear interaction among counter-propagating KAWs (\ref{g.con.WD})
produces an imbalanced turbulence with weakly turbulent omnidirectional
spectra
\begin{equation}
W_{k\perp }\propto k_{\perp }^{-2};  \label{1D.con.WD.WT1}
\end{equation}%
\begin{equation}
W_{k\perp }\propto k_{\perp }^{-1},  \label{1D.con.WD.WT2}
\end{equation}%
created by the direct enstrophy and inverse energy cascades, respectively
(Voitenko, 1998b). The counter-propagating interactions cannot produce steep
observed spectra in the transition range.

At first sight, interaction among counter-propagating KAWs in the weakly
dispersive range is stronger than among co-propagating KAWs: $\gamma
_{k\uparrow \downarrow }^{NL}\gg \gamma _{k\uparrow \uparrow }^{NL}$ for $%
\mu \ll 1$. However, short correlation times among counter-propagating KAWs,
$\tau _{c\uparrow \downarrow }\sim \lambda _{z}/V_{A}$, can reduce their
interaction strength as compared to the co-propagating KAWs that keep in
phase longer time, $\tau _{c\uparrow \uparrow }\sim \mu ^{-2}\lambda
_{z}/V_{A}\gg \tau _{c\uparrow \downarrow }$ for $\mu \ll 1$. In this case,
the dominant (outward-propagating in the solar wind) component of the
imbalanced turbulence will be shaped by the nonlinear interaction among
co-propagating KAWs with omnidirectional spectra (\ref{1D.co.WD.WT3}), as
described above.

The $k_{z}$ spectra of the weak KAW turbulence are determined by the
kinematics of three-wave resonant interactions:
\begin{equation}
W_{k\parallel }\propto k_{z}^{-1/2}.  \label{par.spec.}
\end{equation}%
However, some non-kinematic factors, like finite resonance width, can make
the parallel spectrum significantly different from (\ref{par.spec.}). This
point needs further investigation.

\subsection{Strong KAW turbulence}

In the strongly turbulent regime the nonlinear time scale $\tau
_{k}^{NL}\left( \sim 1/\gamma _{k}^{NL}\right) $ becomes equal to or larger
than the linear one, $\tau _{k}^{L}\left( \sim 1/\delta \omega _{k}\right) $%
, so that the enstrophy should not be conserved any more. In this regime we
find the energy $k_{\perp }$-spectrum from the condition that the energy
flow through any $k_{\perp }$ does not depend on $k_{\perp }$:
\begin{equation}
\varepsilon \sim B_{k}^{2}/\tau _{k}^{NL}=\mathrm{const.}  \label{flux}
\end{equation}%
The nonlinear evolution time for co-propagating KAWs can be estimated as $%
\tau _{k\uparrow \uparrow }^{NL}\simeq 1/\gamma _{k\uparrow \uparrow }^{NL}$%
, where $\gamma _{k\uparrow \uparrow }^{NL}$ is given by (\ref{g.co.WD}).

Then from (\ref{flux}) we find the scaling for the fluctuating magnetic
amplitude $B_{k}\propto k_{\perp }^{-1}$, which results in the
omnidirectional energy spectrum
\begin{equation}
W_{k\perp }\sim \frac{B_{k}^{2}}{k_{\perp }}\propto k_{\perp }^{-3}.
\label{1D.co.WD.ST}
\end{equation}%
Again, one can observe steeper spectra $\propto \left( k_{\perp }^{-3}\div
k_{\perp }^{-4}\right) $ if the strong KAW turbulence is not exactly axially
symmetric around $B_{0}$.

Since $\gamma _{k\uparrow \uparrow }^{NL}$ depends on $k_{z}$ only through $%
B_{k}$, the $k_{z}$-dependence can appear via any functional form with
argument involving some combination of $k_{z}$ and $k_{\perp }$. Additional
assumptions linking $k_{z}$ and $k_{\perp }$, like critical balance
hypothesis, will be studied in another paper.

The strongly turbulent spectra of weakly dispersive counter-propagating KAWs
can be found from (\ref{flux}) with $1/\tau _{k}^{NL}\sim \gamma _{k\uparrow
\downarrow }^{NL}$ given by (\ref{g.con.WD}):
\begin{equation}
W_{k\perp }\sim \frac{B_{k}^{2}}{k_{\perp }}\propto k_{\perp }^{-7/3}
\label{1D.con.WD.ST}
\end{equation}

\section{Strongly dispersive $k_{\perp }\protect\rho _{p}>1$ sub-range of
the KAW turbulence}

In the strongly dispersive sub-range of KAWs $\mu >1$, which was considered
in the literature and named as the "KAW range" (see Schekochihin et al.,
2009, and references therein), the rate of nonlinear interaction among
co-propagating KAWs is (Voitenko, 1998a;b)%
\begin{equation}
\gamma _{k\uparrow \uparrow }^{NL}\simeq 0.3\Omega _{i}\frac{V_{A}}{V_{Ti}}%
\mu ^{2}\frac{B_{k}}{B_{0}}.  \label{g.co.SD}
\end{equation}%
For counter-propagating KAWs, the nonlinear interaction rate is almost the
same,
\[
\gamma _{k\uparrow \downarrow }^{NL}\simeq 0.2\Omega _{i}\frac{V_{A}}{V_{Ti}}%
\mu ^{2}\frac{B_{k}}{B_{0}}.
\]

\subsection{Weak turbulence ($\protect\gamma _{k}^{NL}\ll \protect\omega %
_{k} $)}

The weakly turbulent perpendicular wavenumber spectra of co-propagating
KAWs,
\begin{equation}
B_{k}^{2}\propto k_{\perp }^{-7/2};  \label{2D.co.SD.WT1}
\end{equation}%
\[
B_{k}^{2}\propto k_{\perp }^{-3},
\]%
are created by the direct energy and inverse enstrophy cascades,
respectively (Voitenko, 1998b). Again, the nonlinear rate for the
counter-propagating KAWs can be reduced by shorter linear correlation times
as compared to the co-propagating KAWs. Therefore, the omnidirectional
spectra
\begin{equation}
W_{k\perp }\propto k_{\perp }^{-5/2},  \label{1D.SD.WT}
\end{equation}%
\begin{equation}
W_{k\perp }\propto k_{\perp }^{-2}  \label{1D.SD.WT2}
\end{equation}%
can be formed by strongly dispersive KAWs in the weakly turbulent regime.
Among these, the $\propto k_{\perp }^{-5/2}$ spectrum formed by the direct
energy cascade is preferable. With local deviations from axially symmetry,
one can expect steeper spectra $\propto k_{\perp }^{-2.5}\div k_{\perp
}^{-3} $.

\subsection{Strong turbulence ($\protect\omega _{k}\sim \protect\gamma %
_{k}^{NL}$)}

In the strong turbulence of co-propagating KAWs, the scaling of magnetic
field amplitude $B_{k}$ with $k_{\perp }$ is found from the condition (\ref%
{flux}) where $\tau _{k}^{NL}\simeq 1/\gamma _{k\uparrow \uparrow }^{NL}$
with $\gamma _{k\uparrow \uparrow }^{NL}$ given by (\ref{g.co.SD}):
\[
B_{k}\propto k_{\perp }^{-2/3}.
\]%
This results in the familiar omnidirectional energy spectrum in $k_{\perp }$%
:
\begin{equation}
W_{k\perp }\sim \frac{B_{k}^{2}}{k_{\perp }}\propto k_{\perp }^{-7/3}.
\label{1D.SD.ST}
\end{equation}%
The "parallel" $k_{z}\parallel B_{0}$ spectrum
\begin{equation}
W_{k\parallel }\propto k_{z}^{-2}  \label{1D.SD.ST.par}
\end{equation}%
follows from the critical balance condition.

\section{MHD/kinetic Alfv\'{e}n transition}

\subsection{Spectral kinks}

In the\ Goldreich and Sridhar (1995) MHD model, the AW nonlinear interaction
rate at scale $\lambda _{\perp }$ in the plane $\perp B_{0}$ can be written
as
\begin{equation}
\gamma _{k}^{GS}\simeq \frac{v_{\lambda \perp }}{\lambda _{\perp }}\simeq
\frac{1}{2\pi }k_{\perp }V_{A}\frac{B_{k}}{B_{0}},  \label{GS}
\end{equation}%
where $v_{\lambda \perp }$ is the velocity and $B_{k}$ is the magnetic field
amplitude at the scale $\lambda _{\perp }=2\pi /k_{\perp }$. The
corresponding MHD\ AW\ spectrum $B_{k}^{2}\propto k_{\perp }^{-2/3}$ follows
from the independence of the energy flux through $k$. The omnidirectional
spectrum $W_{k}\sim B_{k}^{2}/k_{\perp }\propto k_{\perp }^{-5/3}$ is seen
by satellites in the\ MHD range as the 1D Doppler frequency spectrum.

As the MHD and weakly dispersive KAW sub-ranges have very different slopes,
the first spectral kink should appear at the wavenumber where their
respective nonlinear interaction rates are equal. Comparing the nonlinear
rates, $\gamma _{k}^{GS}=\gamma _{k\uparrow \downarrow }^{NL}$, we obtain
the spectral kink wavenumber $\mu _{1}\simeq 0.5$, at which the 1D spectrum
should change from $-\left( 3/2\div 5/3\right) $ to $-\left( 3\div 4\right) $%
. The transition wavenumber for the $\gamma _{k}^{GS}=\gamma _{k\uparrow
\uparrow }^{NL}$ transition is practically the same, $\mu _{1}\simeq 0.6$.

However, above estimations did not take into account the weakening of MHD
nonlinear interactions by the dynamic alignment between velocity and
magnetic perturbations (Boldyrev, 2005) and/or by the nonlocal decorrelation
mechanism proposed by Gogoberidze (2007). In general, the interaction rate
can be written as a reduced GS rate (RGS)
\begin{equation}
\gamma _{k}^{RGS}\simeq R_{\lambda \perp }\gamma _{k}^{GS}  \label{g.RGS}
\end{equation}%
with the scale-dependent reducing coefficient $R_{k\perp }$. Both Boldyrev's
and Gogoberidze's phenomenologies give the same scaling for $R_{k\perp }$,%
\[
R_{k\perp }\simeq \frac{v_{\lambda \perp }}{V_{N}}\propto \lambda _{\perp
}^{1/4},
\]%
but with different normalization velocities $V_{N}$, such that the
Boldyrev/Gogoberidze ratio $=v_{L}/V_{A}$, where $v_{L}$ is the velocity
amplitude at the driving scale $L$ (wavenumber $k_{L}$). Having in mind that
the dynamic alignment saturates when approaching small scales, the actual
value of the Gogoberidze coefficient can be larger even in the case $%
v_{L}<V_{A}$. The reduced interaction rate proposed by Gogoberidze can be
written as
\begin{equation}
\gamma _{k}^{RGS}\simeq R_{\lambda \perp }\left( \frac{v_{\lambda \perp }}{%
\lambda _{\perp }}\right) \simeq \left( \frac{k_{\perp }}{k_{L}}\right)
^{-1/4}\left( \frac{1}{2\pi }k_{\perp }V_{A}\frac{B_{k}}{B_{0}}\right) .
\label{g.Giga}
\end{equation}%
Given the typical width of the MHD inertial range in the solar wind $%
k_{b\perp }/k_{L}\sim 10^{3}$, we find that the interaction rate is reduced
considerably in the vicinity of break points, $\gamma _{k}^{RGS}\simeq
0.25\gamma _{k}^{GS}$.

As the nonlocal decorrelation mechanism implies counter-propagating MHD
waves, the counter-propagating KAWs should undergo the same decorrelation.
But co-propagating KAWs do not suffer from such decorrelation, and therefore
we consider here the MHD/kinetic transition dominated by the nonlinear
interaction among co-propagating KAWs. In addition, the co-propagating KAWs
can keep in phase much longer than the counter-propagating KAWs. We
therefore use (\ref{g.co.WD}) for kinetic and (\ref{g.RGS}) for MHD
interaction rate, and estimate the first spectral kink between the shallow
MHD spectra $-\left( 3/2\div 5/3\right) $ and steep weakly dispersive KAW
spectra $-\left( 3\div 5\right) $:
\begin{equation}
\mu _{1}\simeq 0.6\sqrt{R_{k\perp }}.  \label{kink1}
\end{equation}%
With Gogoberidze's rate (\ref{g.Giga}) $\mu _{1}\simeq 0.2$. But one should
bear in mind that there are a number of factors, including a partial
turbulence dissipation, which contribute to $R_{k\perp }$ and can make it
smaller or larger than the Gogoberidze's value.

The second kink point should appear between weekly ($\mu _{p}^{2}\ll 1$) and
strongly ($\mu _{p}^{2}\gg 1$) dispersive regimes of the KAW turbulence at
\begin{equation}
\mu _{2}\gtrsim 1,  \label{kink2}
\end{equation}%
where we allow for a possible building-up of still steeper slope just above $%
\mu =1$ in the cases where the MHD/KAW transition is not yet completed at $%
k_{\perp }=\rho _{p}^{-1}$. The spectrum slope above $\mu _{2},$ is $-\left(
2.5\div 3\right) $, which is significantly shallower than in the weakly
dispersive range.

\subsection{Spectral forms}

The steepness of spectra in the weakly dispersive range depends on what kind
of KAW turbulence picks up the turbulent cascade at $\mu \simeq 0.2$, weak
or strong. If the critical balance condition holds at $\mu \simeq \mu _{1}$,
then the turbulence of weakly dispersive KAWs is strong above $\mu _{1}$. In
this case, strong KAW turbulence develops a steep energy spectrum $\propto
k_{\bot }^{-3}$ in the weakly dispersive sub-range, connecting shallower MHD
($\propto k_{\bot }^{-5/3}$) and strongly dispersive KAW ($\propto k_{\bot
}^{-7/3}$) spectra.

Significantly steeper spectra in both KAW sub-ranges can be produced by the
weak KAW turbulence and by local deviations from the azimuthal symmetry of
the turbulence (up to about $\propto k_{\bot }^{-4.5}$ in the weakly
dispersive sub-range, and $\propto k_{\bot }^{-3}$ in the strongly
dispersive sub-range). Transition to the weak turbulence regime may be
facilitated by the partial wave dissipation via non-adiabatic ion
acceleration/stochastic heating that does not depend on $k_{z}$ but does
depend on $k_{\perp }$ reducing larger amplitudes at $k_{\perp }>k_{\perp
\mathrm{thr}}$. In such a way, the critical balance between linear and
nonlinear time scales is violated in favor of weak turbulent regime. After
that, the weak turbulent cascade of KAWs develops above $k_{\perp \mathrm{thr%
}}$ and establishes steepest KAW spectra.

In both weak and strong turbulence regimes, the resulting spectra have two
kinks, down and up, with the steepest slopes in between them in the
weakly/mildy dispersive sub-range. In general, the "non-dissipative"
scenario is as follows: the turbulence, driven at a large MHD scale $L_{%
\mathrm{dr}}$ ($k_{\mathrm{dr}}=2\pi /L_{\mathrm{dr}}$), develop the
shallowest $\propto k_{\bot }^{-3/2}\div k_{\bot }^{-5/3}$ spectra in the
MHD sub-range $k_{\mathrm{dr}}<k_{\bot }<k_{\bot 1}$, then it proceeds as a
KAW turbulence with steepest $\propto k_{\bot }^{-3}\div k_{\bot }^{-4.5}$
spectra in the weakly dispersive sub-range $k_{\bot 1}<k_{\bot }\lesssim
k_{\bot 2}$, and then above $k_{\bot 2}$ it proceeds as the KAW turbulence
with $\propto k_{\bot }^{-7/3}\div k_{\bot }^{-3}$ spectrum in the strongly
dispersive sub-range $k_{\bot 2}<k_{\bot }\lesssim k_{\bot \mathrm{diss}}$.
This last sub-range may extend to the dissipative wavenumber $k_{\bot
\mathrm{diss}}$ at electron length scale (Sahraoui et al., 2010).

\begin{figure}[]
\vspace*{2mm}
\par
\begin{center}
\includegraphics[width=8.3cm]{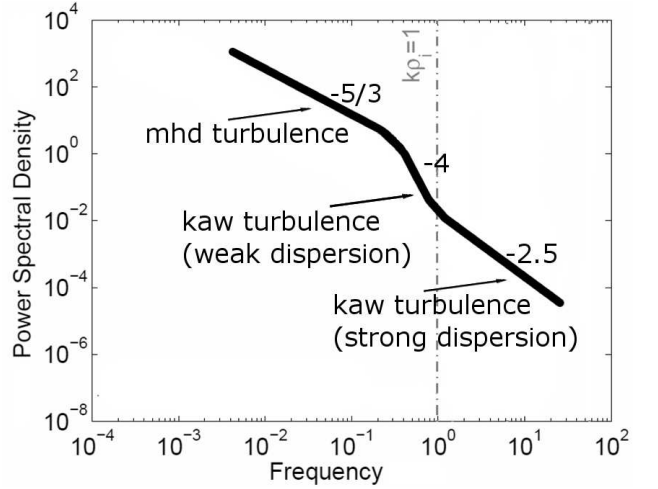}
\end{center}
\caption{Double-kink pattern produced by the MHD/weakly dispersive/strongly
dispersive turbulence transitions. The kinetic KAW spectra are given for the
weakly turbulent regime (in the strongly turbulent regime, spectra -4 and
-2.5 are replaced by -3 and -7/3, respectively)}
\label{fig2}
\end{figure}

Consequently, Alfv\'{e}nic turbulent spectrum in the MHD/kinetic transition
range attains an universal double-kink form (2-k pattern). This 2-k spectral
pattern is shown schematically in Fig. 2 for the case of purely nonlinear
non-dissipative transition. A (variable) slope of the weakly dispersive KAW
spectrum depends on the ratio of turbulent energies cascading in strong
and weak turbulent regimes, which can differ from case to case. The local slope
should in principle lie between -3 and -4. But the shallower >-3 spectra can be
produced by a fraction of the MHD cascade extending above the first kink, and
the steeper <-4 spectra can be produced by azimuthal asymmetry of the turbulence
in the cross-field plane.

A similar 2-k pattern can in principle be produced by the dissipative transition,
suggested by many previous authors, but conditions required for that are rather special.
Namely, the relative dissipation rate (as compared to the nonlinear interaction rate)
should be much stronger in the range $k_{\bot 1}<k_{\bot }<
k_{\bot 2}$ than in the range $k_{\bot }>k_{\bot 2}$.

In any case, the presence of
a high-wavenumber cascade and turbulence at $k_{\bot }>k_{\bot 2}$ imply
the nonlinear transfer and spectral flux in the range $k_{\bot 1}<k_{\bot }\lesssim
k_{\bot 2}$ as well, which means the nonlinear interaction should be taken
into account. In real situations the relative importance of
the effects due to dissipation versus weak turbulence versus strong turbulence
in the MHD/kinetic turbulence transition can be different from case to case.
The 2-k pattern described above can be noticed in many high-resolution
high-frequency Cluster measurements (see examples in papers by Kyiani et
al., 2009; Sahraoui et al., 2010; Chen et al., 2010), and can be also
noticed in some previous measurements where frequency extended to 1 Hz or a
little above (see e.g. Fig. 1 by Smith et al. 2006, showing ACE data).

An actual wavenumber range where super-adiabatic ion acceleration and
related wave damping come into play is also variable. The threshold behavior
suggests that with stronger spectral fluxes it comes into play earlier and
weakens the MHD turbulence. In the cases where the flatness follows the
trends shown in Fig. 3 by Alexandrova et al. (2008) for Cluster data, the
super-adiabatic ion acceleration and partial wave damping may be active well
below the apparent spectral kink.

\section{Dissipation of KAWs}

In this section we introduce several pros and cons concerning basic
dissipations mechanisms for KAWs, but their detailed investigation is
postponed for future.

Wu and Yang (2007) considered self-consistent velocities of minor ion
species in KAW solitons and found them distributed proportionally to the ion
mass-to-charge ratio. However, these velocities cannot be interpreted as
thermal ones increasing temperature because they contribute to the
non-thermal line broadening rather than thermal line width. A non-adiabatic
disconnection from the wave fields is needed for the ions to gain some
energy increase after the wave is passed by. Such process was considered by
Voitenko and Goossens (2004), who shown that KAWs undergo strong
non-adiabatic interaction with ions. This interaction require a certain
threshold-like amplitude/wavelength relation for the dissipation switch-on,
but does not require a long-time stochastic walk for the ions to gain a
significant energy increase.

Chandran et al. (2010) have shown that another process related to
non-adiabaticity - stochastic plasma heating - can absorb up to half of the
turbulent cascade flux at $k_{\perp }\rho _{p}\sim 1$. This result implies
the MHD nonlinear rate, which may be not true at $k_{\perp }\rho _{p}\sim 1$
where the KAW nonlinear interaction is faster and can pass more energy in
the high-$k_{\perp }$ range.

Yet another nonlinear interaction of the broadband Alfv\'{e}nic turbulence
with ions, via nonlinear Landau damping, was studied by Nariyuki et al.
(2010), who shown that the ion heating proceeds both along and across the
background magnetic field and produces asymmetric ion velocity
distributions. On the other hand, because of the quasi-linear platea
formation in velocity distribution functions, classic Landau damping can be
highly reduced in the weakly collisional solar wind (Voitenko and Goossens,
2006; Rudakov et al., 2011).

\subsection{Landau damping}

Parallel components of the KAW electric $E_{z\mathbf{k}}$ and magnetic $B_{z%
\mathbf{k}}$ fields make the KAWs efficient in Cherenkov interaction with
plasma species via kinetic mechanisms of Landau and transit-time damping
that were commonly used in estimations of KAW dissipation. However, these
mechanisms are based on the resonant wave-particle interactions that depend
strongly on the local parallel slopes of particle velocity distributions $%
F_{s}\left( V_{z}\right) $ at parallel velocities $V_{z}$ equal to the wave
phase velocity $\omega _{k}/k_{z}$. In particular, quasi-linear diffusion
reduces resonant slopes and Landau damping (Voitenko 2006):
\begin{equation}
\gamma _{L}=\sum_{s}\gamma _{L}^{M}\left( 1+\frac{\tau _{C}}{\tau _{KAW}}%
\right) ^{-1},  \label{Landau.QL}
\end{equation}%
where $\gamma _{L}^{M}$ is the Maxwellian Landau damping, and $\tau _{KAW}$
and $\tau _{C}$ are the characteristic diffusion times of particles due to
KAWs and Coulomb collisions, respectively. KAWs flatten $F_{s}\left(
V_{z}\right) $, Coulomb collisions restore it back to Maxwellian, and
balance between two results in (\ref{Landau.QL}). Whereas the detailed
analysis of (\ref{Landau.QL}) as function of $k_{\perp }$ is quite complex
(subject to separate study), our estimates, similar to that by Voitenko and
Goossens (2006), show that for typical fluctuation level $W_{f}\sim 10^{-1}$
nT$^{2}$/Hz at $k_{\perp }\rho _{p}\sim 1$ in the solar wind conditions $%
\tau _{C}/\tau _{KAW}\gg 1$ for both electrons and protons, and Landau
damping is thus highly reduced. Therefore, conclusion by Podesta et al
(2009) that the KAW turbulence cannot reach electron scales in the solar
wind, based on the Maxwellian Landau damping, should by reconsidered.

\subsection{Non-adiabatic threshold for turbulent dissipation}

The rate of the non-adiabatic cross-field acceleration of the ions $i$ by
oblique Alfv\'{e}n waves is (Voitenko and Goossens, 2004):
\begin{equation}
\gamma _{\mathrm{n-a}}^{2}=\Omega _{i}^{2}\left[ \frac{V_{A}}{\Omega _{i}}%
\left( \frac{c}{V_{A}}\frac{E_{\perp }}{B_{\perp }}-\frac{V_{iz}}{V_{A}}%
\right) \frac{\partial }{\partial x}\frac{B_{\perp }}{B_{0}}-1\right] ,
\label{g.n-a}
\end{equation}%
where $V_{iz}$ is the parallel ion velocity, $E_{\perp }$ and $B_{\perp }$
are the Alfv\'{e}nic electric and magnetic fluctuation, $\mathbf{E}_{\perp
}\perp \mathbf{B}_{\perp }$.

Using $E_{\perp }/B_{\perp }\simeq V_{A}/c$ in the weakly dispersive range,
and ignoring possible field-aligned streaming of ions, the threshold-like
condition for this kind of wave-particle interaction, $\gamma _{\mathrm{n-a}%
}^{2}>0$, can be written in the form
\begin{equation}
\eta _{k}=k_{\perp }\delta _{p}\frac{B_{k}}{B_{0}}>\nu _{i},  \label{thres1}
\end{equation}%
where $\nu _{i}=\Omega _{i}/\Omega _{p}$ is the threshold value for $\eta
_{k}$ above which the particular ion species $i$ is heated
super-adiabatically.

The condition for ion acceleration apply for any particular ion specie, but
the related wave dissipation depend on all ion species and their parameters,
like species abundances, temperatures, etc. Nevertheless, condition for
efficient wave dissipation can still be written in the form (\ref{thres1})
with a super-adiabaticity $\eta _{k}$ in the left hand side, but with
different threshold $\nu _{w}$ in the right hand side, which is not easy to
find. One can guess that the wave threshold should be close to the
acceleration threshold for the dominant ion species $\nu _{w}\sim \nu _{i}$.
Anyway, even without knowing the exact threshold value $\nu _{w}$, it is
possible to derive several useful scalings that can be tested
observationally. So, for a power law scaling of magnetic amplitudes, $%
B_{k}^{2}\propto k_{\perp }^{-q}$, we obtain the spectral dependence of the
super-adiabaticity $\eta _{k}$:
\begin{equation}
\eta _{k}=\eta _{k1}\left( \frac{k_{\perp }}{k_{\perp 1}}\right) ^{1-q/2},
\label{thres}
\end{equation}%
where $B_{k1}$ is the reference magnetic amplitude at the reference length
scale $\lambda _{\perp 1}=2\pi /k_{\perp 1}$, and $\eta _{k1}$ is the
super-adiabaticity at $k_{\perp }=k_{\perp 1}$. For the sake of convenience
we choose the reference wavenumber equal to wavenumber of the first spectral
kink $k_{\perp 1}$.

Since $q\simeq 2/3$ in the MHD range, super-adiabaticity $\eta _{k}\propto
k_{\perp }^{2/3}$ grows with $k_{\perp }$ as long as $k_{\perp }<k_{\perp 1}$%
. But the situation is reversed in the weakly dispersive KAW range $k_{\perp
}>k_{\perp 1}$, where $q\simeq 3$ and super-adiabaticity decreases with $%
k_{\perp }$ as $\eta _{k}\propto k_{\perp }^{-1/2}$. Such spectral $k_{\perp
}$-dependence of $\eta _{k}$ indicates that the most favorable conditions
for super-adiabatic ion heating and related wave dissipation are achieved in
the vicinity of the first spectral kink, $k_{\perp }\simeq k_{\perp 1}$.
This is shown schematically in Fig. 3, where we used the omnidirectional
spectral representation $W_{k}\propto B_{k}^{2}/k_{\perp }\propto k_{\perp
}^{-p}$ with $p=q+1$. Since $p<2$ ($\simeq 5/3$) in the MHD range, and $p>3$
($\simeq 4$) in the weakly dispersive KAW sub-range, the "threshold"
spectrum
\begin{equation}
W_{\mathrm{thr}}\propto \frac{B_{\mathrm{thr}}^{2}}{k_{\perp }}\propto
k_{\perp }^{-3},  \label{thres2}
\end{equation}%
that follows from the super-adiabatic condition, can fall below the observed
turbulent spectrum $W_{k}$ around the first spectral kink.

\begin{figure}[]
\vspace*{2mm}
\par
\begin{center}
\includegraphics[width=8.3cm]{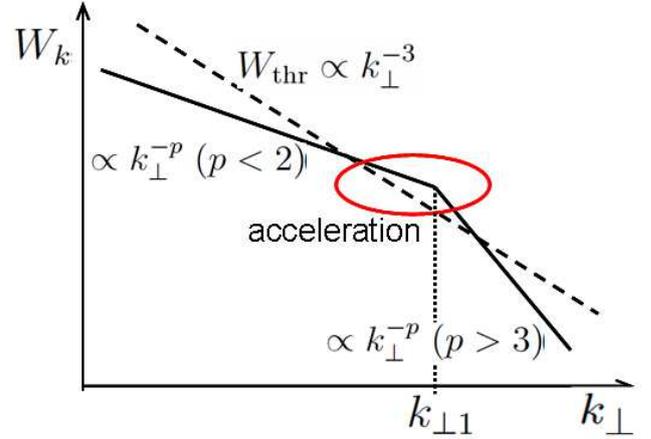}
\end{center}
\caption{Typical Alfv\'{e}nic turbulent spectrum (solid line) in the weakly/mildy dispersive
KAW sub-range and "threshold" turbulent spectrum (dashed line) required for the super-adiabatic ion
acceleration. Super-adiabatic acceleration is possible around the first
spectral kink, where the turbulent spectral power raises above
the threshold spectral power. }
\label{fig3}
\end{figure}

Once the threshold $\eta _{k}=\nu _{i}$ is overcomed in some wavenumber
range for some ion species, the ions enter regime of strong acceleration. In
turn, because of its threshold-like character, the super-adiabatic ion
heating provides a highly selective dissipation mechanism for waves,
affecting only strongest fluctuations with over-threshold amplitudes. In
principle, the ability of turbulence to produce intermittent large-amplitude
fluctuations increases the value of driven parameter (\ref{thres}), where
one should use the spectrum and amplitudes of intermittent fluctuations
instead of the regular turbulent spectrum. The eventual rate of the plasma
heating and turbulence dissipation should follow from the balance between
two processes:\ (i) production of the over-threshold intermittent
fluctuations by the turbulence, and (ii) accommodation of turbulence energy
by accelerated ions and its further redistribution into the bulk plasma.
Helios observations have shown that the flatness (measure of intermittency)
increases with increasing wavenumber (Alexandrova et al., 2008), which
progressively increases the super-adiabaticity parameter above the value
given by (\ref{thres}). Then, at some large enough wavenumber, the level of
intermittent fluctuations can reach the super-adiabatic threshold even if
the regular turbulent level remain below it. .

Dissipation due to super-adiabatic heating/acceleration tends to reduce the
over-threshold fluctuations at every scale to the corresponding threshold
value given by (\ref{thres1}). Then, in accordance to (\ref{thres1}), the
upper bound for the reduced intermittent amplitudes scales as $B_{\mathrm{thr%
}}^{2}\propto k_{\perp }^{-2}$, and since the magnetic power spectrum in
this range has shallower scaling $B_{k}^{2}\propto k_{\perp }^{-2/3}$, the
flatness (and higher order normalized structure functions as well) should
decrease with wavenumber in the MHD range below the first spectral kink.
This can explain another interesting feature, a local decrease of the
flatness in the spacecraft frequency range $0.02\div 0.2$ Hz (which is still
below the apparent spectral kink), found by Alexandrova et al. (2008) using
Cluster data. We suggest that such behavior of the flatness may indicate a
partial dissipation of Alfv\'{e}n waves via super-adiabatic ion acceleration
in the corresponding wavenumber range.

In turn, highly anisotropic ion distributions are produced by
super-adiabatic acceleration (Voitenko and Goossens, 2004), which can drive
anisotropic ion-cyclotron instabilities redistributing energy further. Since
the super-adiabatic ion acceleration is very fast, within a fraction of the
corresponding ion gyroperiod, the quasi-stationary rate of turbulent
dissipation will be determined by the ion-cyclotron instability increment.
The situation is thus more complex here and opposite to that observed in
hydrodynamics, where viscosity washes out smallest amplitudes when
approaching the dissipation range, but large-amplitude fluctuations survive
increasing intermittency. The behavior of the intermittency found by
Alexandrova et al. (2008) is not typical for the linear Landau damping as
well.

After the relative perpendicular/parallel power in the spectrum and the
strength of MHD interaction are reduced, the transition to the weak KAW
turbulence is made possible and leads to the steepest spectra in the weakly
dispersive sub-range. It would be interesting to analyze intermittency by
the Rank-Ordered Multifractal Analysis (ROMA) (see Chang et al., 2010, and
references therein), which would allow to find out if the solar wind
turbulence possesses different fractal properties in the three mentioned
above sub-ranges.

\conclusions[Summary and Discussion]

For the first time, a weakly dispersive sub-range of the KAW turbulence is
distinguished and studied in the context of MHD/kinetic turbulence
transition. We show that the KAW turbulence and its spectra in the weakly
dispersive sub-range differ significantly not only from the conventional MHD
Alfv\'{e}nic turbulence, but also from the strongly dispersive KAW
turbulence. Namely, the nonlinear interaction of weakly dispersive KAWs is
capable to produce steepest spectra $\propto k_{\perp }^{-3}\div k_{\perp
}^{-5}$ in the wavenumber range $k_{\perp 1}<k_{\perp }<\rho _{p}^{-1}$,
connecting shallow MHD spectra $\propto k_{\perp }^{-3/2}\div k_{\perp
}^{-5/3}$ below the first spectral kink, $k_{\perp }<k_{\perp 1}$, and
"intermediate" $\propto k_{\perp }^{-7/3}\div k_{\perp }^{-3}$ spectra of
strongly dispersive KAWs above the second spectral kink, $k_{\perp }>\rho
_{p}^{-1}$.

The universal spectral form - 2-k pattern resulting from such spectral
dynamics in the transition range is shown schematically in Fig. 2.
Turbulent spectra observed recently by the Cluster spacecraft often exhibit
such 2-k pattern in the transition wavenumber range (see for example Fig.1).
It is still not certain what is the role of Landau damping in producing so
steep spectral drop at $k_{\perp }\rho _{p}\lesssim 1$. Any kind of kinetic
dissipation in the weakly collisional solar wind should be self-consistently
saturated at a reduced level by the local platea formation in the velocity
distribution functions of plasma species. At least a quasi-linear theory is
needed to account for the particles' feedback reaction on the energy input
from the waves, and numerous previous estimations based on the Maxwellian
Landau damping should be re-evaluated.

Podesta (2009) reported a significant flattening of the high-frequency
parallel spectra and suggested it may be due to a plasma instability
injecting a fraction of parallel propagating waves. On the other hand, this
flattening can be produced by the transition to the weak KAW turbulence,
possessing (in an ideal case) a very shallow spectrum (\ref{par.spec.}).
However, because of many interfering factors, it is not certain if the
parallel wavenumber spectrum (\ref{par.spec.}) can be realized in the solar
wind. The perpendicular wavenumber spectra are determined by the nonlinear
interaction among perpendicular length scales and are thus quite robust. But
the corresponding parallel wavenumber dynamics and spectra follow the
perpendicular wavenumber dynamics and are often defined from a suitable
functional form linking them to the perpendicular ones. This functional form
may depend on a number of factors, including strength of the turbulence,
partial turbulence dissipation, etc. In the extreme cases of weak and strong
turbulence, the parallel dynamics is fixed, respectively, by the
perpendicular one kinematically (via resonant conditions) and by adjusting
linear and nonlinear time scales (via critical balance condition).

One can expect a high variability of spectral slopes in the weakly
dispersive wavenumber sub-range, resulting from a mixture of several "pure"
spectra that can be produced by KAWs in this range. In addition, our
analysis suggests the super-adiabatic and/or stochastic cross-field
acceleration of the solar wind ions as feasible mechanisms for a partial
KAWs dissipation operating in the vicinity of first spectral kink. Both
these mechanisms share the same non-adiabatic threshold and imply a \textit{%
selective dissipation} of the over-threshold fluctuations with largest
amplitudes. This kind of dissipation reduces high-amplitude intermittent
fluctuations and should therefore produce local decrease of flatness in the
dissipation range. Albeit there are observational indications for such
behavior of flatness (see Fig. 3 by Alexandrova et al., 2008), this point
needs further observational support.

It seems that the synergetic action of selective wave dissipation and weak
turbulence of KAWs influences both the spectral kink positions and the
spectral slopes, making them dependent of the turbulence level. Namely, $%
\eta _{k}$, product of the turbulent amplitude and corresponding wavenumber,
is the parameter facilitating transition to the weak KAW turbulence with its
steeper spectra. As the spectral flux $\sim \eta _{k}^{3}$, the larger
spectral fluxes imply larger $\eta _{k}$, which in turn imply steeper
spectra in the weakly dissipative range. Such a counter-intuitive trend was
found by Smith et al. (2006).

On the other hand, in the vicinity of spectral kinks the non-adiabatic
wave-particle interaction tends to reduce $\eta _{k}$ to a near-threshold
value, which results in the scaling $B_{k1}\sim k_{\perp 1}^{-1}$. This
scaling offer an explanation for observed spectral kink wavenumbers that
were found to be inversely proportional to the fluctuation amplitudes at
spectral kinks (Markovskii et al., 2008).

Contrary to MHD Alfv\'{e}n waves, the dispersion law of KAWs, even weakly
dispersive, is not degenerated with respect to $k_{\perp }$. This makes
possible 3-wave interactions with all 3 waves residing on the KAW branch,
and there is no need in a zero-$k_{\parallel }$, $k_{\perp }\neq 0$ mode
mediating the MHD turbulent cascade. Consequently, an additional spectrum of
the KAW turbulence can be created by the cascading enstrophy (dispersive
part of energy). The energy and the enstrophy cascade in opposite directions
from the injection wavenumber. As the turbulence of KAWs in the solar wind
is driven at largest near-MHD length scales, it naturally proceed to smaller
scales following a direct cascade route. In other circumstances, and with
different positions of the driving scale, one may observe inverse (e.g. Lui
et al., 2008), or dual spectral transport, which is not easy to discriminate
and describe in terms of cascades because of the non-local contaminations
and scale mixing in a real finite-size and high-variable environment (see V%
\"{o}r\"{o}s et al., 2010).

Again, contrary to the MHD AW turbulence, the KAW turbulence does not
require the pre-existing counter-propagating waves for efficient cascading.
Nonlinear interaction among co-propagating KAWs is strong enough to
establish a co-propagating (completely imbalanced) KAW turbulence without
involving the counter-propagating KAWs. If the co-propagating KAW turbulence
develops in some wavenumber range (e.g. at $k_{\perp 1}<k_{\perp }<\rho
_{p}^{-1}$), then the ratio of sunward/anti-sunward Poynting fluxes should
be frozen and remain approximately constant at these wavenumbers. This would
provide another observational benchmark for the KAW turbulence, but we are
not aware about such observations so far.

\begin{acknowledgements}
This work was supported in part by STCE (Solar-Terrestrial Center of
Excellence) under the project "Fundamental science". Some results of this
paper were presented and discussed at the Turbulence and Multifractals
Workshop (8-11 June 2010, Brussels, Belgium).
\end{acknowledgements}


\end{document}